\documentstyle[12pt]{article}
\newcommand{\be}{\begin{equation}}
\newcommand{\nn}{\nonumber}
\newcommand{\bea}{\begin{eqnarray}}
\newcommand{\eea}{\end{eqnarray}}
\newcommand{\ba}{\begin{array}}
\newcommand{\ea}{\end{array}}
\newcommand{\ee}{\end{equation}}

\expandafter\ifx\csname mathbbm\endcsname\relax

\else

\fi \textheight 22cm \textwidth 15cm \topmargin 1mm \oddsidemargin
5mm \evensidemargin 5mm

\newcommand{\beas}{\begin{eqnarray*}}
\newcommand{\eeas}{\end{eqnarray*}}
\newcommand{\bes}{\begin{equation*}}
\newcommand{\ees}{\end{equation*}}

\newcommand{\lf}{\left}
\newcommand{\ri}{\right}
\newcommand{\f}{\frac}

\def\cO{{\cal O}}

\def\tr           {\mbox{\rm tr}\,}

\def\i2           {\mbox{$\frac{i}{2}$}}

\def\al           {\alpha}

\def\bet           {\beta}

\def\tf           {{\tilde F}}
\def\del           {\delta}

\def\ep           {\epsilon}

\def\vep           {\varepsilon}

\def\ga           {\gamma}

\def\Ga           {\Gamma}

\def\la           {\lambda}

\def\om           {\omega}

\def\ph           {\phi}

\def\ps           {\psi}

\def\rh           {\rho}

\def\si           {\sigma}

\def\th{\theta}

\def\pl           {\partial}

\def\ran          {\rangle}
\def\lan          {\langle}

\def\hga{{\hat \gamma}}
\def\we {{\wedge}}

\begin{document}

\begin{titlepage}
\hfill \vbox{
    \halign{#\hfil         \cr
           } 
      }  
\vspace*{20mm}
\begin{center}
{\LARGE \bf{{D-instantons in Klebanov-Witten model}}}\\ 

\vspace*{15mm} \vspace*{1mm} {Ali Imaanpur}

\vspace*{1cm}

{\it Department of Physics, School of Sciences\\ 
Tarbiat Modares University, P.O. Box 14155-4838, Tehran, Iran\\
Email: aimaanpu@theory.ipm.ac.ir}\\
\vspace*{1mm}

\vspace*{1cm}

\end{center}

\begin{abstract}
We study D-instanton solutions in type IIB supergravity on $AdS_5\times T^{1,1}$, which has 
a dual ${\cal N}=1$ $SU(N)\times SU(N)$ super Yang-Mills theory. Apart from ordinary 
D(-1)-brane instantons, we discuss wrapped D1-branes over minimal 2-cycles and derive 
explicit solutions preserving half the supersymmetries. These solutions are 
identified with Yang-Mills instantons which are (anti)self-dual in both gauge group 
factors with instanton charge $(k,k')$. By examining the boundary behaviour of the 
solutions we discuss the coupling to the corresponding dual boundary operators, and identify their 
vacuum expectation values. We also discuss the boundary terms and compute the action for these solutions. 

\end{abstract}

\end{titlepage}

\section{Introduction}
AdS/CFT correspondence has provided us with a duality between type IIB string theory on $AdS_5\times S^5$ and ${\cal N}=4$ SYM theory in four dimensions \cite{MAL}.  
Since the duality relates weak coupling of one theory to the strong coupling of the other, to examine it one needs some quantities to be computed nonperturbatively on both sides of the correspondence. Instantons on the Yang-Mills side, and D-instantons on the supergravity side, being exact solutions of the equations of motion, provide a tool for this purpose. 

The main reasons that allow us to identify instantons with D-instantons -- and also their corresponding contributions in gauge and string theory -- are as follows. First, both of these configurations are exact solutions of the equations of motion: instantons being solutions of Yang-Mills equations on flat space, and D-instantons being solutions of supergravity equations in the background of $AdS_5\times S^5$ supplemented with a 
5-form field strength. Second, the action attains its absolute minimum for these configurations. And finally, they both preserve equal amount of supersymmetries. These two last  properties are especially important in arguing the stability of such solutions when interpolating between weak and strong couplings. 

In this note we look at the model put forth by Klebanov and Witten (KW) aimed at providing AdS/CFT dualities with 
less supersymmetries \cite{KW}. The model is concerned with the dynamics 
of $N$ D3-branes at the tip of a conifold, which, in turn, results in a duality between an ${\cal N}=1$ $SU(N)\times SU(N)$ superconformal field  theory and type IIB string theory on $AdS_5\times T^{1,1}$, where $T^{1,1}$ is the base of the conifold. We will be mainly interested in the supergravity equations of motion in this background, which has the characteristic of having two moduli \cite{KW}. 
The first moduli consists of the constant dilaton and the axion, and the second one arises since the second betti number of $T^{1,1}$ is one. Hence, one can take non-trivial $B$ fields with vanishing field strengths. Generically, adding any extra D-branes to this set up would backreact 
on the geometry, and thus one needs to solve the whole equations of motion from the start. However, there exist  supersymmetric D-instantons which have a zero energy momentum tensor and thus the background 
remains the same as before adding the D-instantons.

As in type IIB supergravity on $AdS_5\times S^5$, one can consider D(-1)-branes in the KW model which source the axion 
and dilaton fields in the bulk. However, since $T^{1,1}$ admits a harmonic 2-form one can wrap (Euclidean) D1-branes, or F-strings, around the dual cycle to get extra point charges in $AdS_5$. Moreover, in this model axion couples to 
the sum of the instanton densities in the corresponding two gauge groups, whereas $C_2$ (for which D1-brane is a source)  couples to the difference 
of these densities. Therefore, D(-1)-branes in the bulk (when $C_2 =0$) correspond to instantons of charge $(k,k)$ on 
the boundary,\footnote{Each $k=\f{1}{16\pi^2}\int \tr(F\we F)$ indicates the instanton number in the corresponding  gauge group.} while D1-branes in the bulk correspond 
to instantons of charge $(k,k')$ with $k\neq k'$. This shows that we need both D(-1)-branes and 
D1-branes to get a complete set of supersymmetric instanton charges on the boundary. 

In the next section, first we look at the supergravity field equations and write down a consistent ansatz to 
reduce them to some coupled scalar field equations on $AdS_5$. In section 3, we look at the supersymmetry of the 
solutions and find that they preserve half the supersymmetries. In section 4, we solve the scalar field 
equations and find explicit solutions. The solutions correspond to D(-1)-branes and D1-branes smeared 
over $T^{1,1}$. Further, we study the behaviour of solutions near the boundary of $AdS_5$ which allows us 
to identify the vacuum expectation values (vev) of the dual operators. In the last section, we write down 
the action in a dual description and derive a lower bound which is saturated by the D-instanton solutions. 
We argue that the bound is set by a boundary term which we then explicitly compute for the solutions we 
have found.

\section{The background and dual operators}
D-instantons in $AdS_5\times S^5$ background are characterized with a non-trivial profile of dilaton 
and axion. In fact, preserving supersymmetry requires
\be
de^{-\ph}=\pm idc\, ,
\ee
where $\ph$ and $c$ are the dilaton and axion fields, respectively. Moreover, with this ansatz D-instantons will have a zero contribution to the energy momentum tensor such that there is no backreaction on the metric. The equation of motion for the dilaton reduces to:
\be
d*d\, e^{\ph}= 0\, ,\label{DMIN}
\ee
which then has an exact solution on $AdS_5\times S^5$ background \cite{KOV, CHU}. 
In this article, we would like to study D-instanton configurations in the Klebanov-Witten model \cite{KW}. The case of 
D(-1)-branes will be similar to what we have in $AdS_5\times S^5$ background, though, in the KW model axion 
at the boundary couples to 
\be
\int \tr\lf(F_1\wedge F_1 +  F_2\wedge F_2\ri)\, ,
\ee
where $F_1$ and $F_2$ are the field strengths taking value in the corresponding gauge group $SU(N)\times SU(N)$.

To start, let us briefly go through the KW model and see how the couplings to the dual operators are arranged. 
The type IIB supergravity background that we are concerned with is characterized by a metric on $EAdS_5\times T^{1,1}$ 
\be
ds^2=\f{L^2}{z^2}\lf(dz ^2 + dx^i dx_i   \ri) + L^2 ds^2_5 \, ,\label{POI}
\ee
where $i=1,\ldots ,4$,  $L^4=4\pi g_s N (\al')^2$, and 
\be
ds^2_5= \f{1}{9}(d\ps + \cos \th_1 d\ph_1 + \cos \th_2 d\ph_2 )^2 + \f{1}{6}\sum_{a=1}^2(d\th_a^2 + \sin^2 \th_a d\ph_a^2)  
\ee
which is the metric on $T^{1,1}=(SU(2)\times SU(2))/U(1)\sim S^2\times S^3$. The five form field strength reads 
\be
F_5 = 4L^4\lf({\rm vol} (AdS_5) + {\rm vol} (T^{1,1})\ri)\, .
\ee
The dilaton and axion are constant, whereas $B= b\, \omega_2$ with a constant $b$, and $\omega_2$ a harmonic 2-form proportional to the volume of $S^2$ in $T^{1,1}$:
\be
\omega_2 = \f{1}{\sqrt {2}}(\sin \th_1 d\th_1\wedge d\ph_1 - \sin \th_2 d\th_2\wedge d\ph_2)\, ,\label{OM2}
\ee
so that 
\be
d\omega_2=d*_5\omega_2=0\, ,
\ee
and
\be
\omega_2\wedge *_5\omega_2 = {\rm vol} (T^{1,1})\, .
\ee

As argued in \cite{KW}, we identify the couplings and the theta angles of the gauge theory with the boundary values of  
the supergravity fields as follows
\bea
&& \f{\th_1}{2\pi} + \f{\th_2}{2\pi} = n +c\, ,\ \ \ \ \ \ \f{4\pi}{g_1^2} + \f{4\pi}{g_2^2}= \f{1}{g_s} \nn \\
&& \f{\th_1}{2\pi} - \f{\th_2}{2\pi} =  n+\int_{S^2}C_2 \, ,\ \ \ \ \ \ \f{4\pi}{g_1^2} - \f{4\pi}{g_2^2}= 
\int_{S^2} B\, , \label{C}
\eea
where $n$ is an integer, and $1$ and $2$ subindices refer to the two $SU(N)$ gauge factors. 
The corresponding gauge theory operators can be clearly identified if we rearrange the 
gauge kinetic and topological terms as follows:  
\bea
\f{1}{g_1^2} \int\tr(F_1\wedge *F_1) + \f{1}{g_2^2}\int\tr( F_2\wedge *F_2)= 
\f{1}{2}\lf(\f{1}{g_1^2} +\f{1}{g_2^2}\ri)\int \cO_\ph + \f{1}{2}\lf(\f{1}{g_1^2} -\f{1}{g_2^2}\ri)\int  \cO_{B}
\, ,\nn \\
 {\th_1} \int\tr(F_1\wedge F_1) + {\th_2}\int\tr( F_2\wedge F_2)= 
\f{1}{2}(\th_1+\th_2)\int \cO_c + \f{1}{2}(\th_1-\th_2)\int  \cO_{C_2}\, ,\ \ \ \ \ \ \ \nn
\eea
where we have defined 
\bea
\cO_{\ph} \equiv   \tr\lf(F_1\wedge *F_1 +  F_2\wedge *F_2\ri)\, ,\label{O11}\\
 \cO_{B} \equiv  \tr\lf(F_1\wedge *F_1 -  F_2\wedge *F_2\ri)\, ,\label{O22}\\
\cO_c \equiv   \tr\lf(F_1\wedge F_1 +  F_2\wedge F_2\ri)\, ,\label{O33}\\
 \cO_{C_2} \equiv  \tr\lf(F_1\wedge F_1 -  F_2\wedge F_2\ri)\label{O44}\, ,
\eea
so that, having in mind the identification in (\ref{C}), we have chosen the subindices to indicate the sources to which 
the operators couple. In particular, this shows that if we turn on a non-trivial $c$ field in the bulk its boundary value acts as a source for $\cO_c$, while $\cO_{C_2}$ couples to the boundary value of $C_2$ field. 
On the other hand, if $C_2=0$ then, as we will argue in Sec. 3.3, the vev of $\cO_{C_2}$ vanishes 
and thus it would correspond to instantons of $(k,k)$ charge at the boundary. We conclude that for having 
instantons of $(k,k')$ charge (with $k\neq k'$) it is necessary to turn on D1-branes.

\section{Smeared D-instantons}
To start our discussion of solutions in the KW model, let us look at the type IIB field equations and see how 
by making suitable ansatz we can reduce them to scalar field equations in $AdS_5$. Next, we consider the 
supersymmetry transformations and find that the ansatz solutions preserve half of the supersymmetries. 

The action of type IIB supergravity in the Einstein frame reads
\bea
&& S=\f{1}{2k_{10}^2}\int d^{10}x {\sqrt g}\, R -\f{1}{4k_{10}^2}\int \lf(d\ph\wedge *d\ph + e^{2\ph} dc\wedge *dc \ri. \nn \\ 
&&\ \ \  \lf. + e^{-\ph} H_{3}\wedge *H_{3} 
+e^{\ph}\tilde{F}_3\wedge *\tilde{F}_3 +\f{1}{2}\tilde{F}_5\wedge *\tilde{F}_5 
+C_4\wedge H_3\wedge F_3 \ri) ,\label{ACT}
\eea
where
\bea
&&F_3=dC_2\, ,\ \ \ F_5=dC_4\, ,\ \ \ H_3=dB\, , \nn \\
&&\tilde{F}_3=F_3-cH_3\, , \ \ \ \ \tilde{F}_5=F_5-C_2\wedge H_3\, ,
\eea
and, 
\be
*\tilde{F}_5 =\tilde{F}_5 \, .
\ee
The type IIB equations of motion derived from the above action read:
\bea
&&d* d\ph =e^{2\ph} dc\wedge *dc -\f{1}{2}e^{-\ph}H\wedge * H+\f{1}{2}e^\ph\tilde{F}_3\wedge *\tilde{F}_3 \nn \\
&&d(e^{2\ph}*dc)=-e^\ph H\wedge *\tilde{F}_3 \nn \\
&&d*(e^{-\ph}H -c e^\ph \tilde{F}_3)=  \, F_3\wedge F_5 \nn \\
&&d*(e^\ph\tilde{F}_3)=\,- H\wedge F_5  \nn \\
&& d*{\tilde F}_5= H\wedge F_3\, .\label{SUG}
\eea

Now, by examining the supersymmetry transformations and the energy momentum tensor we are led to 
consider the following ansatz
\be
de^{-\ph}=\pm idc\, ,\ \ \ \ \ \tilde{F}_3=\pm ie^{-\ph}H\, . \label{BPS0}
\ee
In fact, by looking at the energy momentum tensor
\bea
&& T_{\mu\nu}^{\rm inst}=\f{1}{2}\pl_\mu\ph\pl_\nu\ph-\f{1}{4}g_{\mu\nu}\pl_\rh\ph\pl^\rh\ph +\f{1}{2}
e^{2\ph}\pl_\mu c\pl_\nu c   - \f{1}{4}g_{\mu\nu}e^{2\ph}\pl_\rh c\pl^\rh c  \nn \\
&& +  \f{1}{4}e^{-\ph} H_{\mu\rh\si}H_{\nu}^{\ \rh\si}-\f{1}{24}e^{-\ph}g_{\mu\nu} H_{\rh\si\del}H^{\rh\si\del}
 + \f{1}{4}e^{\ph} \tilde{F}_{\mu\rh\si}\tilde{F}_{\nu}^{\ \rh\si}
 -\f{1}{24}g_{\mu\nu}e^{\ph} \tilde{F}_{\rh\si\del}\tilde{F}^{\rh\si\del}\, ,\nn
\eea
with $\mu, \nu =0, \ldots, 10$, we see that $T_{\mu\nu}^{\rm inst}$ vanishes for the configurations which satisfy (\ref{BPS0}). Therefore, if we could find  
D-instanton solutions with the above ansatz they will have no backreaction on the background metric, and the Einstein 
equations are satisfied in their presence. 

Next, let us see how the field equations are reduced by the above ansatz. For definiteness, however, we choose 
\be
\tilde{F}_3= -ie^{-\ph}H\, ,\label{AN1}
\ee
so that the equations of motion read
\bea
&&d* d\ph =e^{2\ph}dc\wedge *dc -e^{-\ph}H\wedge * H \nn \\
&&d(e^{2\ph}*dc)=iH\wedge *H \nn \\
&&d*(e^{-\ph}H +ic H)=\, F_3\wedge F_5 \nn \\
&& d*H=-\,i H\wedge F_5  \nn \\
&& d*{\tilde F}_5=H\wedge F_3 \, .
\eea
Now, set
\be 
H=dT\wedge \omega_2, \label{AN11}
\ee
where $T$ is a scalar function on $AdS_5$, and $\omega_2$ the harmonic 2-form on $T^{1,1}$ as defined in (\ref{OM2}). The third and fourth equations above imply that we must have 
\be
d(e^{-\ph} +ic)=0\, ,\label{AN2}
\ee 
whereas, the first and the second equations collapse to 
\be
d(e^{\ph}*_5d\ph)= -dT\wedge *_5dT \label{PH}
\ee
together with
\be
d*_5dT=0\, ,\label{F}
\ee
where the subindex $5$ now refers to the $AdS_5$ space. These two last equations are our main concern 
in  this paper, we will solve and discuss them in the subsequent sections.

\subsection{Supersymmetry}
Before discussing the explicit D-instanton solutions, we would like to determine whether the solutions 
satisfying ansatz (\ref{BPS0}) preserve any supersymmetry. For D(-1)-branes it is known that they 
preserve half of the supersymmetries of the background. 

To find a bosonic supersymmetric background the fermionic 
fields are set to zero, then for consistency the supersymmetry variations of these fields must vanish too. 
Hence, in type IIB supergravity we need to check  the supersymmetry transformations of dilatino and gravitino \cite{SCH}. In Euclidean space, for the supersymmetry variation of dilatino we have
\be
\del \la =\f{i}{\tau_2}\Ga^\mu\pl_\mu\tau \vep^*-\f{i}{24}\Ga^{\mu\nu\rh}\vep\, G_{\mu\nu\rh}\, ,
\ee  
\be
\del {\bar \la} =\f{-i}{\tau_2}\Ga^\mu\pl_\mu{\bar \tau} \vep+\f{i}{24}\Ga^{\mu\nu\rh}\vep^* {\bar G}_{\mu\nu\rh}\, ,
\ee 
whereas, the gravitino transformations read
\bea
\del \ps_\mu&=&(\nabla_\mu -\f{i}{2}Q_\mu)\vep+\f{i}{480}\Ga^{\al\bet\ga\rh\si}\Ga_\mu F_{\al\bet\ga\rh\si}\vep 
+\f{1}{96}\lf(\Ga_\mu^{\ \nu\rh\si}G_{\nu\rh\si}-9 \Ga^{\rh\si}G_{\mu\rh\si}\ri)\vep^*\, \nn \\
\del {\bar \ps}_\mu&=&(\nabla_\mu +\f{i}{2}Q_\mu)\vep^*-\f{i}{480}\Ga^{\al\bet\ga\rh\si}\Ga_\mu F_{\al\bet\ga\rh\si}\vep^* 
+\f{1}{96}\lf(\Ga_\mu^{\ \nu\rh\si}{\bar G}_{\nu\rh\si}-9 \Ga^{\rh\si}{\bar G}_{\mu\rh\si}\ri)\vep\, \nn
\eea  
where $\tau=\tau_1+i\tau_2= c+i e^{-\ph}\, ,$ and
\be
Q_\mu =-\f{1}{2}e^\ph \pl_\mu c\, , \ \ \  
G_{\mu\nu\rh}=\f{1}{\sqrt{\tau_2}}(F-\tau H)_{\mu\nu\rh}\, .
\ee
Note that in Euclidean space the spinors and their complex conjugates are in different representations and 
thus are treated as independent fields (parameters). 
 
Now for the background to be supersymmetric we choose $\vep^*=0$. Setting the dilatino variation to zero 
implies
\be
\pl_\mu{\bar \tau}=0\, ,\ \ \ \  \Ga^{\mu\nu\rh}\ep\, G_{\mu\nu\rh}=0\, ,\label{TA} 
\ee
whereas, setting the gravitino variations to zero yields
\bea
&&(\nabla_\mu -\f{i}{2}Q_\mu)\vep+\f{i}{480}\Ga^{\al\bet\ga\rh\si}\Ga_\mu F_{\al\bet\ga\rh\si}\vep =0\, , \label{DM} \\
&&\ \ \ \ \lf(\Ga_\mu^{\ \nu\rh\si}{\bar G}_{\nu\rh\si}-9 \Ga^{\rh\si}{\bar G}_{\mu\rh\si}\ri)\vep=0 \, .\label{GB}
\eea
The ansatz in the previous section, (\ref{AN1}) and (\ref{AN2}), already satisfy the following equations
\be
\pl_\mu{\bar \tau}=0\, ,\ \ \ \  {\bar G}_{\mu\nu\rh}=0\, ,\label{TG} 
\ee
so that the first equation of (\ref{TA}) and eq. (\ref{GB}) are satisfied. Therefore, for being supersymmetric, we further need to check the second equation of (\ref{TA}), i.e.,
\be
\Ga^{\mu\nu\rh}\vep\, G_{\mu\nu\rh}=0\, .\label{EXTRA}
\ee

To show this, we use the following vielbein basis for the metric on $T^{1,1}$:
\bea
e^i&=&\f{L}{\sqrt 6}d\th_i\, , \ \ \ \ e^{\hat i}=\f{L}{\sqrt 6}\sin \th_i d\ph_{ i}\, , \ \ \ { i}=1,2\nn \\
e^3&=&\f{L}{3} (d\ps + \cos \th_1 d\ph_1 +  \cos \th_2 d\ph_2)\, , \ \ \ \nn
\eea
The non-vanishing components of the Riemann tensor are then
\bea
&&R^{1\hat {1}2 \hat {2}}=-2\, ,\ \ \ \ \ R^{1\hat {1} 1\hat {1}}=R^{2\hat {2} 2\hat {2}}=3\\
&&R^{3131}= R^{3\hat {1} 3\hat {1}}= R^{3232}=R^{3\hat {2} 3\hat {2}} = R^{1\hat {2} \hat {1}2}=R^{12 \hat {2}\hat {1}}=1 \, ,
\eea
from which the non-vanishing components of the Weyl tensor can be derived
\bea
&&C^{1\hat {1}\hat {2}2}=C^{1\hat {1} 1\hat {1}}=C^{2\hat {2} 2\hat {2}}=2\\
&&C^{1\hat {2}1\hat {2}}= C^{\hat {1}2 \hat {1}2}= C^{1212}=C^{\hat {1}\hat {2}\hat {1} \hat {2}} = C^{1\hat {2} 2\hat {1}}=C^{12 \hat {1}\hat {2}}=-1\, .
\eea
In this basis we have
\be
\omega_2 = \f{6}{\sqrt{2}L^2}(e^1\we e^{\hat {1}}-  e^2\we e^{\hat{2}})\, .
\ee

Note that before introducing the D-instantons, the background preserves 8 supercharges 
determined by the integrability condition along $T^{1,1}$:\footnote{For the Gamma matrices we take $\Ga^n= \si_1\otimes \hga^n\otimes 1$, for $n = 1, \ldots, 5$, 
and $\Ga^a= \si_2\otimes 1\otimes \ga^a$, for $a=6, \ldots, 10$, with $ \vep= \left(
\begin{array}{l}
1 \\
0	
\end{array}\right)\otimes \ep \otimes \eta$.}
\be
C^{abcd}\ga_{cd}\, \eta =0\, ,
\ee
where $\eta$ indicates a constant spinor on $T^{1,1}$. For $a=1, b=\hat{1}$, this implies
\be
(\ga_{ 1\hat {1}}+\ga_{\hat {2}2})\eta =0\, .\label{INT}
\ee
Going back to (\ref{EXTRA}) we have
\bea
\Ga^{\mu\nu\rh}\vep\, G_{\mu\nu\rh}&=&\f{1}{\sqrt{\tau_2}}\Ga^{\mu\nu\rh}\vep\, (F-\tau H)_{\mu\nu\rh}\ =\ \f{-2i}{\sqrt{\tau_2}}e^{-\ph}\Ga^{\mu\nu\rh}\vep\, H_{\mu\nu\rh}\nn \\
&=&\f{-12i}{\sqrt{\tau_2}}e^{-\ph} \pl_n T\,\left(
\begin{array}{l}
0 \\
1	
\end{array}\right) \otimes \hga^n \ep\otimes (\ga^{1\hat{1}} \om_{1\hat{1}}+\ga^{{2} \hat{2}}\om_{{2} \hat{2}})\eta\nn\\
&=& \f{-72i}{L^2\sqrt{2\tau_2}}e^{-\ph}\pl_n T\, \left(
\begin{array}{l}
0 \\
1	
\end{array}\right) \otimes \hga^n \ep\otimes (\ga_{1\hat{1}}-\ga_{{2} \hat{2}})\eta =0\, ,
\eea
where we have used the ansatz (\ref{AN1}), (\ref{AN11}), and the last line follows because of (\ref{INT}). 
At the end, there remains to check eq. (\ref{DM}). However, this equation is the same that appears when 
we have only D(-1)-branes, which are known to be half supersymmetric. 
So we conclude that our D-instanton solutions preserve 1/2 supersymmetries.

\subsection{Solutions}
Let us now turn to solving the last two equations in Sec. 3. Eq. (\ref{F}), with 
a delta function singularity on the right hand side (here the location of D-instanton)  
in fact defines the Green function of massless scalars in $AdS_5$. The general case of massive 
scalar propagators have been discussed and derived in terms of hypergeometric functions in \cite{FRO, BUR}. 
However, for massless scalars, the solutions are easier to find. 

To begin with, let us write eq. (\ref{F}) in the Poincare coordinates, (\ref{POI}), 
and set $L=1$ for convenience  
\be
z^3\pl_z\lf(\f{1}{z^3} \pl_z T \ri) + \pl_i\pl^i T=0\, .\label{z}
\ee
This equation is the same as eq. (\ref{DMIN}) for D(-1)-branes, it can be conveniently solved by first defining 
\be 
\xi=\f{2 z z_1}{z^2+z_1^2+(\vec{x}-\vec{x_1})^2}\, ,
\ee
in terms of which the geodesic distance is usually expressed. If we write eq. (\ref{z}) in terms of $\xi$, 
it reduces to an ordinary differential equation in variable $\xi$:
\be
4 \xi^2(1-\xi^2) T''(\xi^2) - 2(2+3\xi^2) T'(\xi^2)=0\, .
\ee
The solution is the  D1-branes instanton (smeared along $T^{1,1}$) in $AdS_5$ space
\be
T(\xi^2) = a_1  + b_1 \lf(\f{3\xi^2-2}{(1-\xi^2)^{3/2}}\ri)\, ,\label{T}
\ee
which has a delta function singularity at $\xi^2=1$, the location of the D-instanton 
\be
z=z_1 \, ,\ \ \ \ \ \ \vec{x}=\vec{x_1}\, .
\ee
$a_1$ and $b_1$ are two constant parameters and the boundary value of $T$, when $z\to 0$, is $T_0=a_1 -2b_1$. 

Now, let us discuss dilaton equation (\ref{PH}). Substituting solution (\ref{T}) for $T$ on the 
right hand side of (\ref{PH}), it reduces to 
\be
4 \xi^2(1-\xi^2) (e^\ph)'' - 2(2+3\xi^2) (e^\ph)' +  \f{9b_1^2\, \xi^6}{(1-\xi^2)^4}=0\, ,
\ee
where again prime indicates the derivative with respect to $\xi^2$. This equation can be integrated to get the  particular solution
\be
e^\ph_s = - \f{b_1^2\, (3\xi^2 -2)^2}{2 (1-\xi^2)^3}\, .
\ee
To this we can add the solution of the homogeneous equation to get the general solution for the dilaton:
\be
e^\ph = {a_0} + b_0\,\lf( \f{3\tilde{\xi}^2-2}{(1-\tilde{\xi}^2)^{3/2}}\ri) - \f{b_1^2\, 
(3\xi^2 -2)^2}{2 (1-\xi^2)^3}\, ,\label{SOL}
\ee
where, to keep track of the different moduli, the solution of the homogeneous equation is now written 
in terms of  
\be
\tilde{\xi} = \f{2 z z_0}{z^2+z_0^2+(\vec{x}-\vec{x_0})^2}\, .
\ee
Here, $a_0$ and $b_0$ are two constant parameters so that $e^\ph \to e^{\ph_0} = a_0-2b_0-2b_1^2$ at the boundary.

\subsection{Near boundary behaviour}
Having a scalar field of mass $m$ in $AdS_{d+1}$, we can examine its boundary behaviour. 
According to the AdS/CFT correspondence \cite{W98}, the boundary value of the scalar field acts 
as a source for the dual operator of dimension $\Delta$ 
\be
\Delta= \f{1}{2}(d+\sqrt{d^2+ 4m^2})\, .
\ee 
Furthermore, the subleading term in fact determines the vacuum expectation value of the dual operator \cite{BALA, SKE}. More explicitly, near the boundary we have
\be
\ph (z, {\vec x})\ \sim\ z^{d-\Delta}\, \ph_0({\vec x})  +  z^{\Delta}\, \ph_1({\vec x})\, ,\label{SCA}
\ee
so that $\ph_0({\vec x})$ acts as a source of the dual operator, while $\ph_1({\vec x})$ determines 
its expectation value. 

The scalar fields we are concerned with have zero masses, and hence should couple to dimension 4 operators at the boundary, which we discussed in Sec. 2.    
Now that we have explicit exact solutions of these scalars we can look at the boundary behaviour and read off 
the vacuum expectation values of the dual operators. When $z\to 0$, for the dilaton we find
\be
e^\ph \ \sim \ e^{\ph_0}- 44\, b_0 \f{ z^4 z_0^4}{(z_0^2+(\vec{x}-\vec{x_0})^2)^4}+ 168\, b_1^2  \f{ z^4 z_1^4}{(z_1^2+(\vec{x}-\vec{x_1})^2)^4}\, ,\label{dil}
\ee
whereas, $T$ behaves as 
\be
T \ \sim \ T_0  -44 b_1 \f{ z^4 z_1^4}{(z_1^2+(\vec{x}-\vec{x_1})^2)^4}\, .\label{til}
\ee
Comparing the above equations with (\ref{SCA}), we indeed see that they correspond to dimension 4 boundary operators.  The coefficients of $z^4$ terms in (\ref{dil}) and  (\ref{til}) determine the vev of $\cO_\ph$ and $\cO_B$ 
(as defined in (\ref{O11}) and (\ref{O22})), respectively. 

Now, recall from (\ref{BPS0}) that  
\be
d(e^{-\ph} + ic)=0\, ,\label{BPS00}
\ee
while (\ref{AN11}) implies
\be
{F}_3=dC_2= -id((e^{-\ph} +ic)B)= -id((e^{-\ph} +ic)T\, \om)\, , \label{AN111}
\ee
therefore, we can obtain the corresponding solutions for $c$ and $C_2$, i.e., $c\sim e^{-\ph}$, and $C_2 \sim T\, \om$. Further, using the above recipe the coefficients of the subleading terms in (\ref{dil}) and (\ref{til}) determine the vevs: 
\be
\lan \cO_c(\vec{x}) \ran \sim  \f{ 44\, e^{-2\ph_0} b_0\, z_0^4}{(z_0^2+(\vec{x}-\vec{x_0})^2)^4}-  
\f{168\, e^{-2\ph_0}b_1^2\,  z_1^4}{(z_1^2+(\vec{x}-\vec{x_1})^2)^4} \label{O1}
\ee
and,
\be
 \lan \cO_{C_2}(\vec{x})\ran \sim  \f{b_1\, z_1^4}{(z_1^2+(\vec{x}-\vec{x_1})^2)^4}\, ,\label{O2}
\ee
where $\cO_c$ and  $\cO_{C_2}$ were defined in (\ref{O33}) and (\ref{O44}).

As expected from the gauge theory side, the vev of the operators $\cO_c$ and $\cO_{C_2}$ are proportional to the 
Yang-Mills instanton densities. Furthermore, (\ref{BPS00}) and (\ref{AN111}) now imply
\bea
\lan \cO_{{\ph}}(\vec{x}) \ran = \lan \cO_{c}(\vec{x}) \ran \, ,\\
\ \ \ \ \ \ \lan \cO_{B}(\vec{x})\ran = \lan \cO_{C_2}(\vec{x})\ran \, ,
\eea
hence we conclude that
\bea
\lan \tr(F_1\wedge *F_1) \ran = \lan \tr(F_1\wedge F_1) \ran \, ,\\
\lan \tr(F_2\wedge *F_2) \ran = \lan \tr(F_2\wedge F_2)\ran \, .
\eea
This observation, as in the case of D(-1)-branes, further supports the identification 
of D-instantons with Yang-Mills instantons of the boundary theory, i.e., $F_1 =*F_1$ and $F_2 =*F_2$. 
In this way, the location (moduli) of the D-instanton solutions, e.g., $(z_0, \vec{x}_0)$, is identified with the scale and location of Yang-Mills instantons. Had we chosen the upper sign in (\ref{BPS0}), we would have obtained D-instantons corresponding to $F_1 =-*F_1$ and $F_2 =-*F_2$. Both of these configurations in gauge theory are known to preserve half the supersymmetries. 
On the contrary, we know that instantons of opposite charges, i.e., $F_1 =*F_1$ and $F_2 =-*F_2$, break supersymmetry, 
and, in fact, it is difficult to construct the dual brane configurations in the bulk.

Note that, if $b_0 \neq 0$, and $b_1=0$, namely if we have only D(-1)-branes, then $\lan \cO_{C_2} \ran =0$ and thus 
it would correspond to having instantons of charge $(k,k)$ on the boundary sitting on top of each other 
at $\vec{x}=\vec{x_0}$ with size $z_0$. On the other hand, when $b_1 \neq 0$, and $b_0=0$, i.e., having 
only D1-branes, then $\lan \cO_c \ran$ and  $\lan \cO_{C_2} \ran \neq 0$ and therefore we have instantons of charge 
$(k, k')$, with $k\neq k'$, on the 
boundary. They are all located at $\vec{x}=\vec{x_1}$ with size $z_1$.
The more general case of $b_0,\,  b_1 \neq 0$, is simply the sum of the above instantons at two different 
locations $\vec{x}=\vec{x_0}$, and $\vec{x}=\vec{x_1}$. From (\ref{O1}), we see that if we take 
$(z_0, {\vec x}_0) = (z_1, {\vec x}_1)$ and let $b_0= 42 b_1^2/11$ then we have $\lan \cO_c \ran =0$, which 
would correspond on the boundary with two instantons of opposite charges sitting on top of each other.  


\section{Action and the dual description}
To motivate the boundary terms and hence the action of D-instantons, it is useful to rewrite the action 
in terms of dual fields. This has a twofold benefit of getting the boundary term as well as deriving the 
saturating bounds. The bounds will be the same as those we get from supersymmetry \cite{GUT}. 
So, let us assume that the Chern-Simons term 
is vanishing (as is the case for the solutions we have found), and start with the following action:
\bea
&&S=-\f{1}{4k_{10}^2}\int \lf( d\ph\we *d\ph + e^{-\ph} H\we *H + e^{-2\ph} \tf_9 \we *\tf_9 + 
e^{-\ph} F_7 \we *F_7 \ri.\nn \\
&& \lf.\ \ \ \ \ \ \ \ \ \ \ \ \ \ \ \ \ \ \  -2 c\, (d\tf_9 -F_7\we H) -2\, C_2\we dF_7\ri)\, ,\label{ACT2}
\eea
which upon integration over $c$ and $C_2$ yields the {\em dual} description
\be
S=-\f{1}{4k_{10}^2}\int \lf( d\ph\we *d\ph + e^{-\ph} H\we *H +e^{-2\ph} \tf_9 \we *\tf_9 + 
e^{-\ph} F_7 \we *F_7 \ri)\, , \label{MIN}
\ee
supplemented with the Bianchi identities
\be
\tf_9=dC_8+C_6\we H\, ,\ \ \ \ \ F_7=dC_6\, .\label{BIA}
\ee

Further, let us define the Euclidean action by $S=i S_E$, and write (\ref{MIN}) as\footnote{For $\om$ an $r$-form on an $m$-dimensional manifold, we have $*^2 \om =(-1)^{r(m-r)}\, \om$ for a Riemannian metric, 
and $*^2 \om =(-1)^{1+r(m-r)}\, \om$ when the metric is Lorentzian.}
\bea
&&S_E =-\f{1}{4k_{10}^2}\int \lf( (d\ph-e^{-\ph}*\tf_9)\we * (d\ph-e^{-\ph}*\tf_9) + 
e^{-\ph}(H-*F_7)\we * (H-*F_7) \ri. \nn \\
&& \ \ \ \ \ \ \ \ \ \ \ \ \ \ \ \ \ \ \ \ \ \lf.-2\, e^{-\ph} d\ph\we \tf_9 
-2\, e^{-\ph} H\we F_7 \ri)\, . \label{EUC}
\eea
Using the Bianchi identities, (\ref{BIA}), the last two terms combine to a boundary term
\bea
&& -2 \int (e^{-\ph} d\ph\we \tf_9 + e^{-\ph} H\we F_7 ) \\ 
&& = 2 \int (de^{-\ph} \we (dC_8+C_6\we H) + e^{-\ph} F_7\we H ) =2\int d(e^{-\ph}\tf_9)\label{ACT1}
\eea
therefore, since the first two terms in (\ref{EUC}) are semi positive-definite, we get a lower bound on the Euclidean action
\be
S_E\,  \geq\,  -\f{1}{2k_{10}^2} \int d(e^{-\ph}\tf_9)\, ,\label{ACT3}
\ee
and the bound is saturated whenever
\be
d\ph= e^{-\ph}*\tf_9 \, , \ \ \ \ \ H=*F_7\, .\label{BPS}
\ee
By reversing the signs in (\ref{EUC}), we could saturate the bound with 
\be
d\ph= -e^{-\ph}*\tf_9 \, , \ \ \ \ \ H=-*F_7\, ,
\ee
for which the boundary term flips its sign. However, for a configuration with opposite signs; i.e.,  $d\ph= -e^{-\ph}*\tf_9 \, , \  H=*F_7\, ,$ 
we do not get a lower bound set by a {\em boundary term}. 

Next, we wish to write the action in a dual description. First, let us write action (\ref{ACT2}) in Euclidean space and vary with respect 
to $\tf_9$ and $F_7$:
\[
\del S_E =\f{-1}{2k_{10}^2}\!\int\! \lf( e^{-2\ph} \del\tf_9 \we *\tf_9 + e^{-\ph} \del F_7 \we *F_7 
  + i c\, (d\del \tf_9 -\del F_7\we H) +i \, C_2\we d\del F_7\ri) \,  
\]
where $i$ comes from rotating to Euclidean signature. Setting $\del S_E=0$, we get
\be
\tf_9= i e^{2\ph}*dc\, , \ \ \ \ \ \ \ F_7=i e^{\ph} * \tf_3\, .\label{BPS1}
\ee
Plugging this back into the action gives back the original action (\ref{ACT}), now accompanied by the boundary terms:
\bea
&&S_E=-\f{1}{4k_{10}^2}\int \lf(  d\ph\we *d\ph + e^{2\ph} dc\we *dc +e^{-\ph} H\we *H + e^{\ph} \tf_3 \we *\tf_3 \ri. \\ 
&&\ \ \ \ \ \ \ \ \ \ \ \ \ \ \ \ \ \ \ \ \lf. -2d(c\, e^{2\ph}*dc + e^{\ph} C_2\we *\tf_3)\ri)\, .\label{ACTD}
\eea
We can now see that equations (\ref{BPS}) for which the bound is saturated are in fact configurations (\ref{BPS0}) that 
preserve supersymmetry. Using (\ref{BPS}) and (\ref{BPS1}), we have
\bea
&& d\ph =e^{-\ph}*\tf_9 = -i e^\ph dc \\
&& H=*F_7 = -i e^\ph \tf_3 \, ,
\eea
which are the same equations as (\ref{BPS0}). Further, using the above equations we can see that the boundary terms in both descriptions, i.e.,  
(\ref{ACT1}) and (\ref{ACTD}), coincide. 

Having obtained the lower bound on the action, let us now compute it for the (saturated) solutions that we 
found in the Sec. 3. From (\ref{ACT3}), we have 
\be
S_E = -\f{1}{2k_{10}^2} \int_M d(e^{-\ph}\tf_9) = -\f{1}{2k_{10}^2} \int_M d*d\ph = -\f{1}{2k_{10}^2} \int_{\pl M} *d\ph
\ee
where boundary $\pl M$ consists of a part at $z\to 0$, and two small spheres around the D(-1)-brane at 
$(z_0, {\vec x}_0)$, and D1-brane at $(z_1, {\vec x}_1)$. Taking the exterior derivative of dilaton solution (\ref{SOL}), doing the Hodge dual and then taking the limit $z\to 0$ we obtain
\be
*d\ph |_{z\to 0} = e^{-\ph_0} \lf( \f{48 b_0\, z_0^4}{(z_0^2+(\vec{x}-\vec{x_0})^2)^4} +  
\f{96 b_1^2\, z_1^4}{(z_1^2+(\vec{x}-\vec{x_1})^2)^4} \ri)\, d^4x \we\, {\rm vol} (T^{1,1})\, .
\ee
Finally, the integral over the 4d space and $T^{1,1}$ results to
\be
S_E = \f{128}{27 k_{10}^2} e^{-\ph_0} \pi^5 (b_0+2 b_1^2) \, ,
\ee
where we used
\be
\int {\rm vol} (T^{1,1}) =\f{16 }{27}\pi^3\ ,{\rm and}\ \ \ \ \int \f{d^4x}{(z_0^2 + (\vec{x}-\vec{x_0})^2)^4}=\f{\pi^2}{6 z_0^4}\, ,
\ee
and the fact that the integral over the two small spheres around the D-instantons, when their size 
is shrinking to zero, vanishes.

\section{Conclusions}
In this article, we derived explicit exact solutions of smeared D-instantons in the KW model. 
The D-instantons consist of D(-1)-branes together with D1-branes. Following \cite{KW}, 
we argued that in this model D(-1)-branes source  the sum of the instanton densities in the gauge 
theory, whereas  D1-branes provide a source to the difference of these densities. We saw that with a suitable ansatz 
the equations of motion can be reduced to a coupled scalar field equations, which we then solved and 
examined its boundary behaviour.  This allowed us to determine the vev of the corresponding dual 
operators in the gauge theory. In particular, we observed that to get and identify instantons of charge 
$(k,k')$, with $k\neq k'$, on the boundary we need to include D1-branes in the bulk. We further examined the 
supersymmetry of the solutions and found that, like D(-1)-branes, they preserve half the supersymmetries. 
We also discussed the dual description of the D-instantons and described the dual action which then 
enabled us to write down the boundary terms and find the saturating bounds. At the end, we computed the 
action of these D-instanton solutions.

\end{document}